# Assembling ensembling: a typology for approaches across disciplines

Amanda Bleichrodt[1,2, ¶], Sadie J. Ryan[3,2, ¶*], Lydia Bourouiba[4,2], Gerardo Chowell[5,2], Eric T. Lofgren[6,2], J. Michael Reed[7,2], Nina H. Fefferman[8,2*]

[1] Department of Public Health Sciences, Clemson University, Clemson, SC

[2] US NSF Center for Analysis and Prediction of Pandemic Expansion (APPEX), University of Tennessee, Knoxville, 37996, USA

[3] Department of Geography and Emerging Pathogens Institute, University of Florida, Gainesville, FL 32611 USA

[4] The Fluid Dynamics of Disease Transmission Laboratory, Massachusetts Institute of Technology, Cambridge, Massachusetts 02139, USA

[5] Department of Population Health Sciences, Georgia State University, School of Public Health, Atlanta, GA, USA

[6] Paul G. Allen School for Global Health, Washington State University, Pullman, WA, USA

[7] Department of Biology, Tufts University, Medford, MA 02155 USA

[8] Department of Ecology and Evolutionary Biology, University of Tennessee, Knoxville, TN, 37996-3140, USA

¶These authors contributed equally to this work.

*Co-corresponding author

Email: sjryan@ufl.edu, nina.h.fefferman@gmail.com

## Abstract

When discussing model ensembling or ensemble modeling, a term arises across numerous disciplines, what is meant by it can vary drastically. The very meaning of 'ensemble' - a collection together - conjures different ideas even within disciplines when approaching phenomena. For example, one might think of a set of descriptions of a phenomenon in the world, perhaps a time series or a snapshot of multivariate space, and perhaps that set is comprised of data-independent descriptions, or perhaps it is quite intentionally fit *to* data, or even a suite of data sets with a common theme or intention. Recently, ensemble models have appeared widely across applications, for disease forecasting, environmental suitability modeling, and more. In this piece, we present a typology of the scope of potential perspectives across disciplines to disambiguate terms, concepts, and processes associated with 'ensembles' and 'ensembling'. We do not provide an exhaustive review nor do we recommend that all disciplines must adopt a common suite of terms, but instead focus on facilitating communication, awareness, identification of gaps, and adoption of tools to avoid independent efforts to reinvent the wheel across disciplines. To anchor our discussion, we provide a Shiny App to contain the typology, with a living collection, or compendium, of example publications about ensembles.

## Introduction

The concept of an ensemble has become commonplace in various academic disciplines to describe a wide diversity of phenomena, such as time series data, snapshots of multivariate space, data-independent descriptions, a collection of quantitative approaches, and even data sharing common themes and intentions. A general definition of *ensemble* is a group of elements producing a single effect (https://www.merriam-webster.com/dictionary/ensemble). When multidisciplinary teams coalesce around cross-cutting questions – e.g. can we predict the next dengue outbreak? [1] or what has driven wetland loss in Hawaii since human settlement [2] – we see increasing uses of "ensemble approaches". However, the nuances of defining, naming, and applying ensembling concepts become apparent when comparing literature within and across disciplines. This can lead to confusion in the meaning and application of methods and frameworks in multidisciplinary or transdisciplinary collaborations, creating barriers to communication and impeding progression. Semantic barriers can prevent successful integration across disciplines, so we encourage consideration of the ways terms are used. To that end, we present a typology for disambiguating ensemble and ensembling and provide a R-Shiny App dashboard to illustrate this, using a sample of the literature as examples.

The number of publications and research areas contributing to the body of work using ensembles has increased consistently over at least two decades (Fig 1). The diversity of usage complicates navigating the literature on ensemble-based analysis, and varying terms and applications of "ensemble" may be utilized, even within the same area of study.

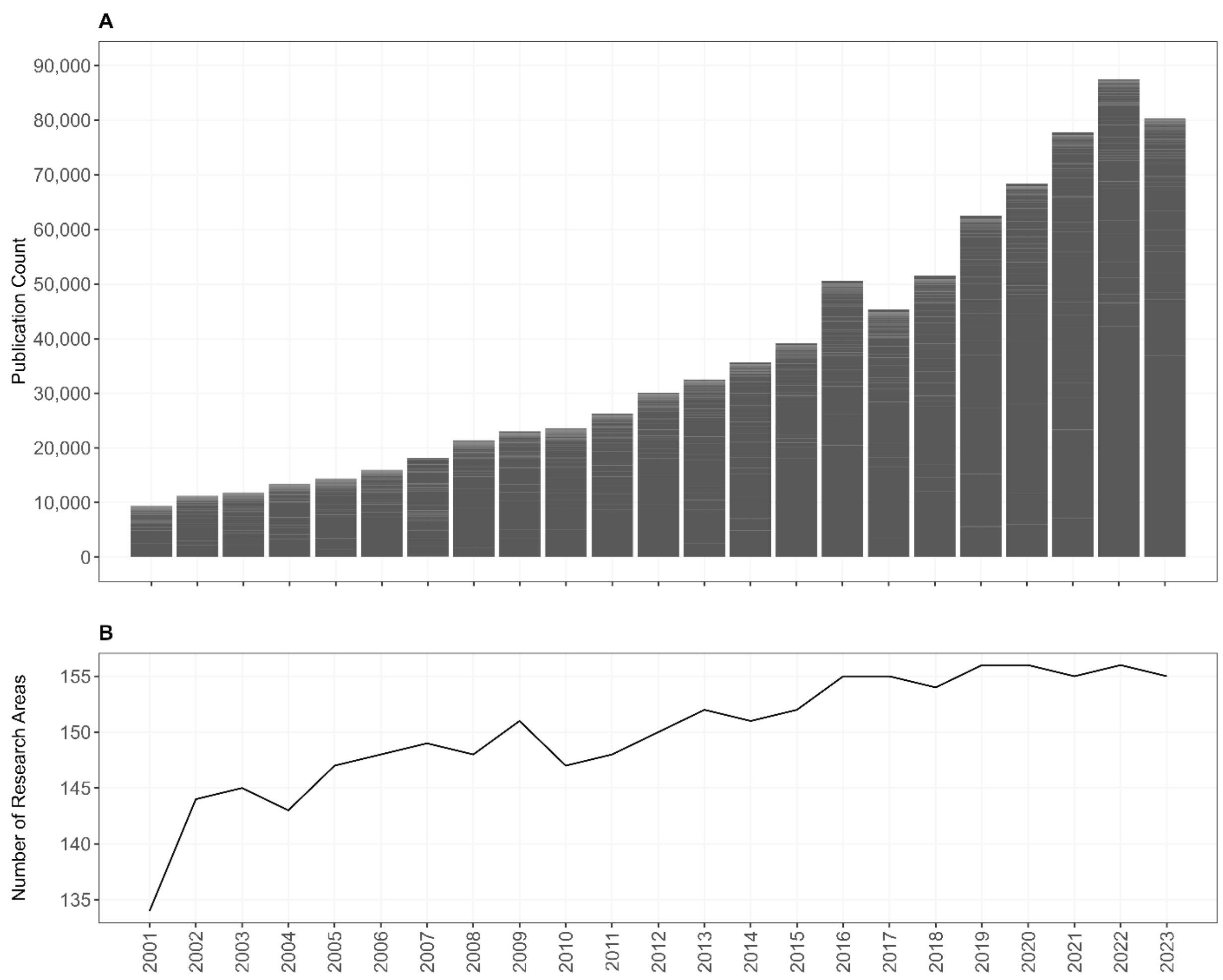

**Fig 1. The frequency of publications including the word "ensemble" and the number of research areas contributing publications each year.** Panel A illustrates the frequency of publications employing the keyword "ensemble" from 2001 through 2023. Panel B shows the number of research areas, as defined by Web of Science, contributing publications employing the keyword “ensemble” each year (search run by AB from Georgia State University; data derived from Clarivate™ (Web of Science) ©Clarivate 2023).

While the application of ensemble-related processes may differ within and across disciplines, the foundational definition of an ensemble generally holds. We posit that nuances arise in defining (1) the items being assembled into a single object, (2) the procedure used to form the ensemble, and (3) the end product of the ensemble process. For example, individuals primarily working with epidemic forecasting may associate the term "ensemble" with combining the predictive power of multiple models to reach one comprehensive forecast [3,4]. In contrast, individuals working regularly with large amounts of data may describe data fusion, the process of harmonizing a large amount of data and variables to allow for a singular analysis as an ensemble process [5–7]. Nevertheless, both examples fall under the commonly used definitions of “ensemble” (as either a noun or a verb), while the terms used to describe the process and the elements being assembled differ.

Given the growing use of ensemble processes across disciplines and the complexity of the associated lexicon, exploring the nature of "ensembling" terminology will facilitate the navigation of existing and future literature that use, and will build on, these concepts. Attempting to provide a comprehensive typology of ensemble phenomena across all existing applications is neither practical nor desirable, as the techniques, disciplines, and lexicons involved are likely to continue to expand. Thus, our primary objective is to provide a unifying map of the diversity of concepts relating to ensembles so that different fields may discover and build on each other’s techniques. A living compendium, prototyped in an R-Shiny app for sorting and defining studies employing ensemble, ensembles, ensembling, and so on, accompanies this brief dive into an expanding literature to facilitate ongoing concept map structuring of the corpus (Fig 2).

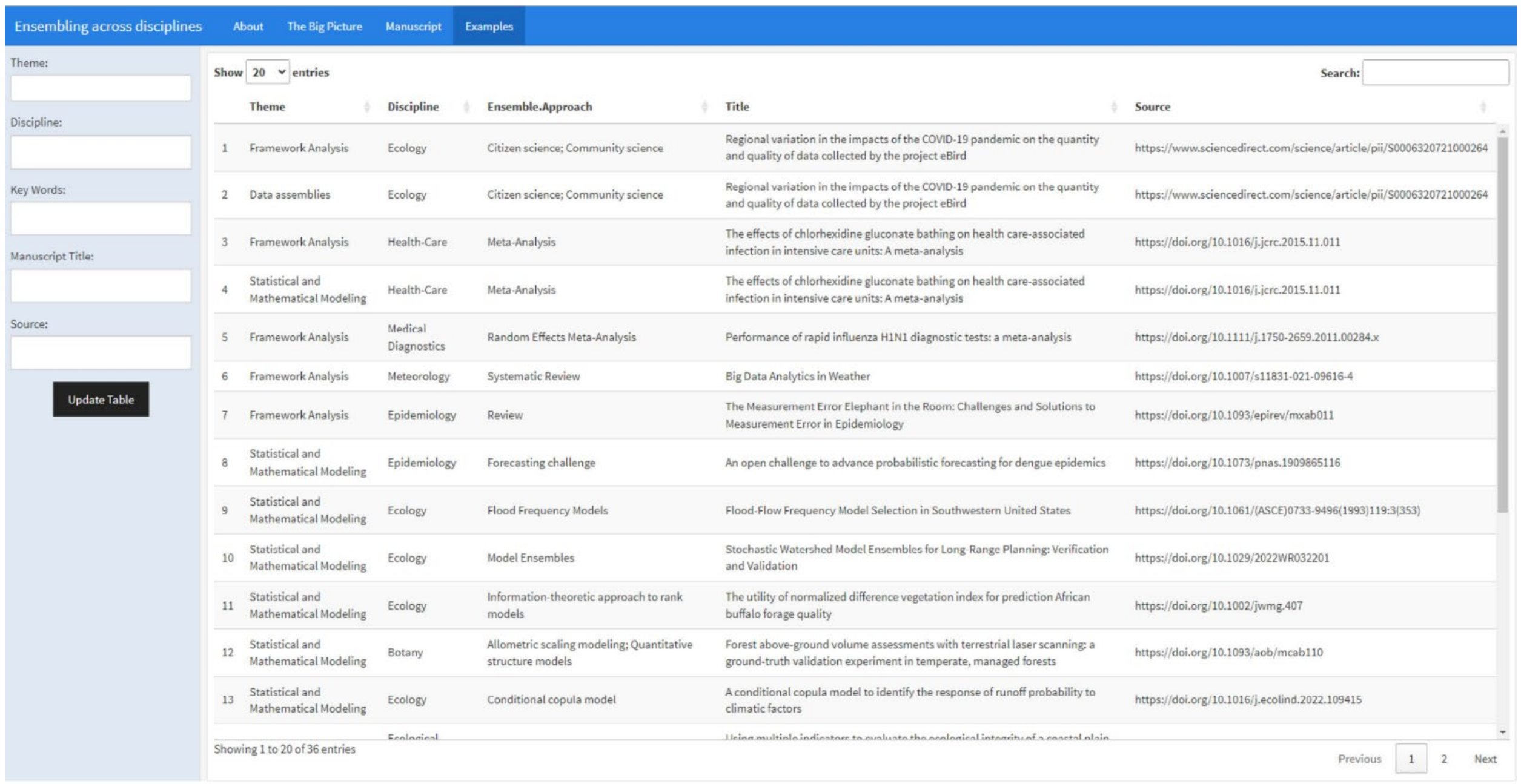

| | Theme | Discipline | Ensemble.Approach | Title | Source |
|---|---|---|---|---|---|
| 1 | Framework Analysis | Ecology | Citizen science; Community science | Regional variation in the impacts of the COVID-19 pandemic on the quantity and quality of data collected by the project eBird | https://www.sciencedirect.com/science/article/pii/S0006320721000264 |
| 2 | Data assemblies | Ecology | Citizen science; Community science | Regional variation in the impacts of the COVID-19 pandemic on the quantity and quality of data collected by the project eBird | https://www.sciencedirect.com/science/article/pii/S0006320721000264 |
| 3 | Framework Analysis | Health-Care | Meta-Analysis | The effects of chlorhexidine gluconate bathing on health care-associated infection in intensive care units: A meta-analysis | https://doi.org/10.1016/j.jcrc.2015.11.011 |
| 4 | Statistical and Mathematical Modeling | Health-Care | Meta-Analysis | The effects of chlorhexidine gluconate bathing on health care-associated infection in intensive care units: A meta-analysis | https://doi.org/10.1016/j.jcrc.2015.11.011 |
| 5 | Framework Analysis | Medical Diagnostics | Random Effects Meta-Analysis | Performance of rapid influenza H1N1 diagnostic tests: a meta-analysis | https://doi.org/10.1111/j.1750-2659.2011.00284.x |
| 6 | Framework Analysis | Meteorology | Systematic Review | Big Data Analytics in Weather | https://doi.org/10.1007/s11831-021-09616-4 |
| 7 | Framework Analysis | Epidemiology | Review | The Measurement Error Elephant in the Room: Challenges and Solutions to Measurement Error in Epidemiology | https://doi.org/10.1093/epirev/mxab011 |
| 8 | Statistical and Mathematical Modeling | Epidemiology | Forecasting challenge | An open challenge to advance probabilistic forecasting for dengue epidemics | https://doi.org/10.1073/pnas.1909865116 |
| 9 | Statistical and Mathematical Modeling | Ecology | Flood Frequency Models | Flood-Flow Frequency Model Selection in Southwestern United States | https://doi.org/10.1061/(ASCE)0733-9496(1993)119:3(353) |
| 10 | Statistical and Mathematical Modeling | Ecology | Model Ensembles | Stochastic Watershed Model Ensembles for Long-Range Planning: Verification and Validation | https://doi.org/10.1029/2022WR032201 |
| 11 | Statistical and Mathematical Modeling | Ecology | Information-theoretic approach to rank models | The utility of normalized difference vegetation index for prediction African buffalo forage quality | https://doi.org/10.1002/jwmg.407 |
| 12 | Statistical and Mathematical Modeling | Botany | Allometric scaling modeling; Quantitative structure models | Forest above-ground volume assessments with terrestrial laser scanning: a ground-truth validation experiment in temperate, managed forests | https://doi.org/10.1093/aob/mcab110 |
| 13 | Statistical and Mathematical Modeling | Ecology | Conditional copula model | A conditional copula model to identify the response of runoff probability to climatic factors | https://doi.org/10.1016/j.ecolind.2022.109415 |

**Fig 2. The living compendium.** This figure shows a snapshot of an R-Shiny app, which was created within R for sorting and defining studies within this typology (https://collabdashboard.shinyapps.io/Shiny-Dashboard/)

Our discussion of ensemble processes outlines three themes our multidisciplinary author team believes best capture underlying types of “ensemble” found in the literature: (1) assemblies of data, (2) framework analysis, and (3) mathematical and statistical modeling. For each theme, we first define common terms used throughout the themed section and then discuss their relations to the foundational definition of an ensemble. We also discuss the nuances of ensemble processes as applied to each of the three themes using examples. Finally, we provide an annotated typology of our concept of the scope of ensembles and ensemble modeling, with examples of real-world ensemble-based literature across disciplines, as a sample in our R-Shiny dashboard (Table 1).

**Table 1.** Examples from a variety of disciplines where the concept of 'ensemble' or 'ensembling' is used; this sample was used to initially populate our dashboard.

| Title | Theme | Discipline | Ensemble Approach | |
|---|---|---|---|---|
| Regional variation in the impacts of the COVID-19 pandemic on the quantity and quality of data collected by the project eBird | Data assemblies; framework analysis | Ecology | Citizen science; Community science | [8] |
| Machine learning methods for predicting progression from mild cognitive impairment to Alzheimer's disease dementia: a systematic review | Ensemble learning | Biometrics | Machine learning for epidemiological application | [9] |
| Random Forests for Classification in Ecology | Statistical and mathematical modeling | Ecology | Random Forest | [10] |
| A practical introduction to Random Forests for genetic association studies in ecology and evolution | Statistical and mathematical modeling | Molecular Ecology | Random Forest | [11] |
| Assessing the accuracy and stability of variable selection methods for random forest modeling in ecology | Statistical and mathematical modeling | Environmental Monitoring and Assessment | Random Forest Modeling | [12] |
| A Tale of Two "Forests": Random Forest Machine Learning Aids Tropical Forest Carbon Mapping | Statistical and mathematical modeling | Ecology | Random Forest Machine Learning | [13] |
| The effects of chlorhexidine gluconate bathing on health care-associated infection in intensive care units: A meta-analysis | Framework Analysis | Healthcare | Meta-Analysis | [14] |
| Performance of rapid influenza H1N1 diagnostic tests: a meta-analysis | Framework Analysis; Statistical and Mathematical Modeling | Medical Diagnostics | Random Effects Meta-Analysis | [15] |
| Big Data Analytics in Weather Forecasting: A Systematic Review | Framework Analysis | Meteorology | Systematic Review | [16] |

| The Measurement Error Elephant in the Room: Challenges and Solutions to Measurement Error in Epidemiology | Framework Analysis | Epidemiology | Review | [17] |
|---|---|---|---|---|
| Doubly robust estimation of causal effects | Statistical and mathematical modeling | Epidemiology | Doubly Robust estimation | [18] |
| Targeted Maximum Likelihood Estimation for Causal Inference in Observational Studies | Statistical and mathematical modeling | Epidemiology | Doubly Robust Maximum-likelihood-based approach | [19] |
| An open challenge to advance probabilistic forecasting for dengue epidemics | Statistical and mathematical modeling | Epidemiology | Forecasting challenge | [1] |
| Flood-Flow Frequency Model Selection in Southwestern United States | Statistical and mathematical modeling | Ecology | Flood Frequency Models | [20] |
| Stochastic Watershed Model Ensembles for Long-Range Planning: Verification and Validation | Statistical and mathematical modeling | Ecology | Model Ensembles | [21] |
| Wetland Loss in Hawai'i Since Human Settlement | Statistical and mathematical modeling | Ecology | Topographic wetness index | [2] |
| The utility of normalized difference vegetation index for prediction African buffalo forage quality | Statistical and mathematical modeling | Ecology | Information-theoretic approach to rank models | [22] |
| Ecology Cues, Gestation Length, and Birth Timing in African Buffalo (*Syncerus caffer*) | Framework analysis; Statistical and Mathematical modeling | Ecology | Data ensembling; Information theoretic approach to rank models | [23] |
| Forest above-ground volume assessments with terrestrial laser scanning: a ground-truth validation experiment in temperate, managed forests | Statistical and mathematical modeling | Botany | Allometric scaling modeling; Quantitative structure models | [24] |

| A conditional copula model to identify the response of runoff probability to climatic factors | Statistical and mathematical modeling | Ecology | Conditional copula model | [25] |
|---|---|---|---|---|
| The Stream Quality Index: A multi-indicator tool for enhancing environmental management | Framework analysis; Statistical and Mathematical modeling | Environment al and Sustainability Indicators | Stream Quality Index (SQI) | [26] |
| Using multiple indicators to evaluate the ecological integrity of a coastal plain stream system | Framework Analysis | Ecological Indicators | Multiple indicators | [27] |
| Determining habitat quality for species that demonstrate dynamic habitat selection | Statistical and mathematical modeling | Ecology | Wading bird distribution and evaluation models (WADEM) | [28] |
| Drivers of distributions and niches of North American cold-adapted amphibians: evaluating both climate and land use | Statistical and mathematical modeling | Ecology | Combination models of climate and land use | [29] |
| Integrating ensemble species distribution modelling and statistical phylogeography to info projections of climate change impacts on species distributions | Statistical and Mathematical Modeling | Ecology | Ensemble species distribution modeling and statistical phylogeography | [30] |
| Assessing abiotic correlations of an indicator species with sympatric riparian birds in a threatened submontane river-forest system using joint species modeling | Statistical and Mathematical Modeling | Ecology | Joint species modelling | [31] |
| Multiple regression and inference in ecology and conservation biology: further comments on identifying important predictor variables | Statistical and Mathematical Modeling | Ecology | Multiple regression | [32] |
| Ensemble Forecasts of Coronavirus Disease 2019 (COVID-19) in the U.S. | Statistical and Mathematical Modeling | Epidemiology | Ensemble | [33] |
| Retrospective evaluation of short-term forecast performance of ensemble sub-epidemic | Statistical and Mathematical Modeling | Epidemiology | Ensemble Frameworks | [34] |

| frameworks and other time-series models: The 2022-2023 mpox outbreak across multiple geographical scales, July 14th, 2022, through February 26th, 2023 | | | | |
|---|---|---|---|---|
| A review of statistical methods for the evaluation of aquatic habitat suitability for instream flow assessment | Framework analysis; Statistical and Mathematical modeling | River Research | Review; Habitat statistical models | [35] |
| PCA - A Powerful Method for Analyzing Ecological Niches | Statistical and Mathematical modeling | Ecology | Principal Component Analysis (PCA) | [36] |
| Multiscale soil and vegetation patchiness along a gradient of herbivore impact in a semi-arid grazing system in West Africa | Statistical and Mathematical modeling | Ecology | Habitat modeling | [37] |
| Transcending scale dependence in identifying habitat with resource selection functions | Statistical and mathematical modeling | Ecology | Multi-scale resource selection modeling | [38] |
| Uncovering multiscale effects of aridity and biotic interactions on the functional structure of Mediterranean shrubland | Framework Analysis; Statistical and mathematical modeling | Ecology | Habitat filtering; Niche Differentiation; Functional trait-based and multiscale approach | [39] |
| Multi-scale assessment of human-induced changes to Amazonian instream habitats | Statistical and Mathematical modeling | Landscape Ecology | Random forest regression trees | [40] |
| A Multi-Scale Test of the Forage Maturation Hypothesis in a Partially Migratory Ungulate Population | Statistical and Mathematical modeling | Ecology | Habitat modeling | [41] |
| Metamodels for Transdisciplinary Analysis of Wildlife Population Dynamics | Statistical and Mathematical Modeling | Ecology | Metamodels | [42] |
| Joint analysis of demography and selection in population | Statistical and Mathematical Modeling | Genetics | Joint analysis | [43] |
| genetics: where do we stand and where could we go? | | | | |
| Biological ensemble modeling to evaluate potential futures of living marine resources | Statistical and Mathematical Modeling | Ecology | Ensemble modeling | [44] |
| Testing whether ensemble modelling is advantageous for maximizing predictive performance of species distribution models | Statistical and Mathematical Modeling | Ecology | Ensemble modeling | [45] |
| Ensemble forecasting of species distributions | Statistical and Mathematical Modeling | Ecology | Ensemble forecasting frameworks | [46] |
| Multi-source data fusion for aspect-level sentiment classification | Framework Analysis; Statistical and Mathematical Modeling | Ecology | Data Fusion | [5] |
| A modified flexible spatiotemporal data fusion model | Statistical and mathematical modeling? | Ecology | Data Fusion | [47] |
| A systematic review of data fusion techniques for optimized structural health monitoring | Framework Analysis | Health Monitoring | Data Fusion; Systematic Review | [6] |
| Multi-Source Data Fusion Improves Time-Series Phenotype Accuracy in Maize under a Field High-Throughput Phenotyping Platform | Framework Analysis | Ecology | Multi-Source Data Fusion | [7] |
| Data fusion approaches for structural health monitoring and system identification: Past, present, and future | Framework Analysis | Health Monitoring | Data Fusion | [48] |
| Mixture of Regression Models With Varying Mixing Proportions: A Semiparametric Approach | Statistical and Mathematical Modeling | Statistics | Semiparametric mixtures of regression models | [49] |
| Discrete-Time Hazard Regression Models with Hidden Heterogeneity: The | Statistical and Mathematical Modeling | Sociology | Mixed Poisson Regression Approach | [50] |

| Semiparametric Mixed Poisson Regression Approach | | | | |
|---|---|---|---|---|
| Ensemble modelling in descriptive epidemiology: burden of disease estimation | Statistical and Mathematical Modeling | Epidemiology | Ensemble modeling | [51] |

**Theme 1: Assemblies of Data**

Across academic fields of study, researchers often combine individual observable elements (i.e., data), to form a more extensive collection of such elements (i.e., a data set) or a set of related collections (i.e., data repository) for later use, [8,52–54]. In this study, elements refer to individual observations (data) that can be combined to reach the level of a data set; a data set is a collection of elements (data) and a data repository is a collection of data sets (Fig 3A).

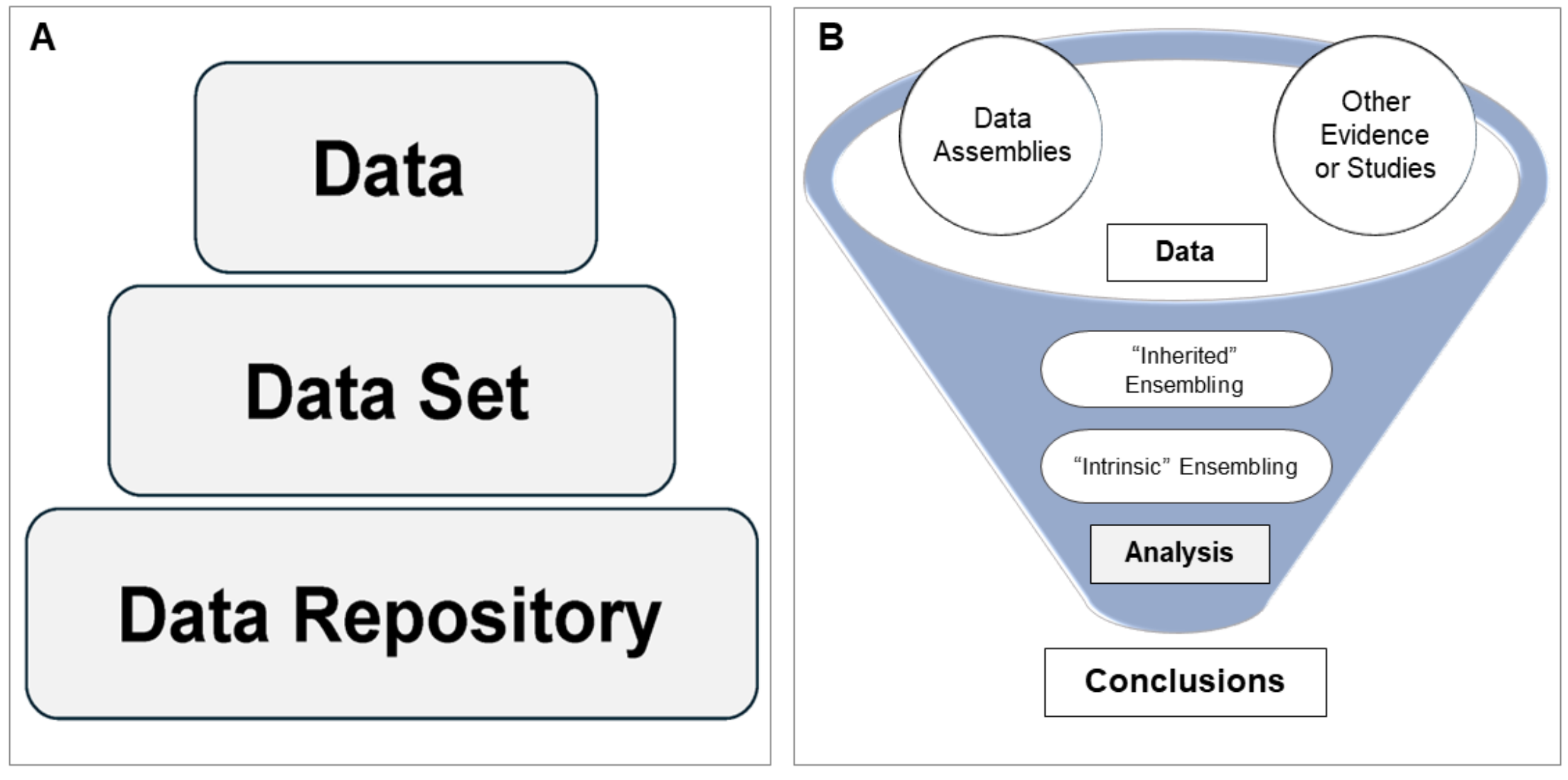

**Fig 3. A visualization of the three encompassing levels of data assemblies and the ensembling nature of the framework analysis process.** Panel A illustrates the encompassing nature of the levels of data assemblies (i.e., data, data set, and data repository). As can be seen in the figure, data sets include data elements, whereas data repositories contain both combinations of data and data sets. Panel B visualizes the ensembling nature of the framework analysis process and possible complexities that exist in the literature.

### *Defining data assemblies and their collection*

Data is a non-specific term often used to describe individual pieces of factual, observable information collected to improve our understanding of a system, e.g., via statistical or mathematical analysis. As the term encompasses a broad range of subject matter (i.e., any observable, measurable, or reliably inferable information), complexities can arise in both its composition and collection. For this paper, each *element of data* will refer to a single observation. In contrast, a *data set* will be used to describe a collection of data, and a *data repository* refers to a collection of data sets. In each level shown in Fig 3A, analysts combine multiple units of the type from the previous box, such as elements of data or data sets, into one entity (i.e., a data set or data repository, respectively). Thus, the act of combining information at varying levels falls under the purview of an ensemble [26,27,55]. However, how the data are collected or measured, their source(s), and motivation for forming the ensemble may differ.

### *Complexities of ensembling within data assemblies*

When looking at data sets and data repository ensembling, the underlying building blocks (Fig 3A) often come from multiple sources while still assisting a single collected outcome. While *source* is a vague term, its meaning in ensembling is data set and repository-specific. For example, if working with weather data, a researcher may collect the daily rainfall measurement from many rain stations and combine each datum into one data set [56]. Thus, the underlying building blocks (elements of data; i.e., rainfall values) that are assembled to produce the data set (i.e., the ensemble of those elements) present information about the same topic but come from different sources (i.e., rain stations). In data repository ensembling, *source* may refer to many teams combining data sets into one collection. Thus, in this example, *source* now refers to a specific research team. It may be that the act of ensembling depends on how the data or data sets that contribute to the data set or repository (respectively) were themselves collected or assembled; different techniques may be used depending on whether method of collection of the units now being assembled were homogeneous or heterogeneous across all sources.

Continuing with the rainfall example, each station may collect rain using different instruments, though each still provides observations of the rainfall for the data set. The act of ensembling these elements may then require a normalization step to assemble them coherently into a usable data set (harmonization) that provides consistent information about the same thing (i.e., rainfall) from the different sources. Therefore, the process meets the criteria of ensembling regardless of whether we utilize the same methods during the datum or data set collection process. When data sets are combined into a data repository, similar principles hold; the composition and collection methods used for both data sets and data repositories do not change the general application of the ensemble definition to the process of forming either [57].

Critically, what we here refer to as data ensembling is often described using other terminology across disciplines. The above-described nuances of collection and composition require different techniques and structures to allow analysis of the assembled information to provide insight and understanding. Different fields have developed protocols and tools to enable these analyses and, due both to siloed efforts and the independent development of language to describe the methods, these parallel (and sometimes redundant) paths of progress in data assemblies remain largely isolated from, and uninformed by, each other. Perhaps even more problematic is that the parallel disciplines that tackle these challenges may pick similar (or identical) terms to describe different techniques. As no comprehensive list of terms to classify data assemblies currently exists in the literature, and such collections are critical to exploring data assemblies-related literature effectively, we provide multiple illustrative examples in Table 1. We intend this to serve as the seed for an online living cross-referencing resource (see Fig 2), fed by ongoing contributions from disparate communities, rather than an already-exhaustive list.

**Theme 2: Framework Analysis**

The formation of data assemblies is a critical step in the exploratory process. It is rarely, however, the end in itself. Researchers must employ additional methods to derive meaningful conclusions from data assemblies. While this often comes in the form of a statistical or mathematical model (discussed below), other systematic methods lend themselves to being conceptualized as ensembling methods. We term these methods “framework analyses”, recognizing a conceptual link to qualitative data analysis methods [17,58,59].

***Defining framework analysis***

In the existing literature, framework analysis is often presented as a comprehensive, systematic perspective and protocol, applied to qualitative data assemblies with the primary goal of pattern identification and data exploration [60]. We extend this definition beyond the traditionally included qualitative data methods to include quantitative data methods that use predefined, systematic methods to analyze data. Framework analysis can be thought of as an ensemble in one of two ways: 1) “Inherited” ensembling, where the combination of different data sources into a data assembly, as described above, is the primary way by which multiple data sources are incorporated into a single analysis, or 2) “Intrinsic” ensembling, where this integration takes place at the analytical stage (Fig 3B). Finally, while statistical and mathematical modeling approaches fall under this category, the breadth and nuance of these methods dictate addressing these in a separate section (see below).

***Complexities of ensembling within framework analysis***

To continue the rainfall example, an analysis of a data assembly of rainfall from multiple weather stations over time, summarizing trends, mapping them, etc. would be an inherited ensembling process [16]. The analysis itself does not involve additional drawing of disparate data or analyses beyond what is contained in the data assembly that goes into the analysis and is functionally similar to an analysis conducted on a single data source. While not often viewed as ensembling, this sort of practice is potentially its most common form in many fields, as many descriptive studies, machine learning, spatial methods, etc. fall into this category [9,13,61,62]. Framework analysis that can be regarded as intrinsic ensembling consists of analytical frameworks that draw upon multiple data sources as part of the analysis itself, without necessarily incorporating them into a data assembly first (or if they do, it is at a purely functional level for the use of analytical software). The meta-analysis of other studies falls within this category [63], as do systematic reviews, and slightly less quantitative methods that rely on an expert or group of experts leveraging multiple disparate pieces of evidence or studies to inform their own internal models from which to draw conclusions or make recommendations [14,15]. This brings methods such as the "Modified Delphi Method" [60], and processes for developing many clinical guidelines under the umbrella of ensemble methods, albeit potentially absent a rigorously specified model of how a conclusion is drawn from the ensemble.

Regardless of whether the ensembling is inherited from the data assembly (or assemblies) used in the analysis or intrinsic to the analysis itself, both of these types of framework analyses meet our more general definition. Ultimately, multiple elements (i.e., data assemblies) are combined to reach a single effect - in this case, conclusions about the data assemblies or outcome of interest. Although we present multiple examples highlighting the complex nature of our version of framework analysis across different academic disciplines, our list of possible scenarios is non-exhaustive. Therefore, the above sections function as a generalized description of the common nuances within our definition of framework analysis to facilitate the exploration of the related literature (see examples in Table 1).

### Theme 3: Statistical and Mathematical Modeling

The most intricate tools for integrating, analyzing and interpreting data assemblies involve the design, implementation, and application of models. Such models are built by researchers and analysts in many fields and range from general use to explore any data conforming to format requirements to those specifically tailored to address disciplinary questions of interest. As discussed above, modeling is a particular type of framework analysis; however, beyond direct analyses of data assemblies alone, it can also involve manipulation and transformation of data assemblies (both within and across scales) and culminates in a produced projection of system behavior to predict data in (as yet) unobserved scenarios (whether into the future or into uncharted areas of the parameter space) [10,12,21,64].

#### *Defining Statistical and Mathematical Modeling*

Modeling techniques can be either primarily statistical or mathematical in nature. Statistical models derive from interrogating the data assemblies themselves to discover patterns and exploit those patterns to make predictions based on the expected continuation of those patterns [65]. For example, positive correlational patterns dictate that an increase in one variable indicates an increase in another, allowing prediction of either one from the other. Note that variable B may or may not depend causally on variable A in order to observe a positive correlation, but in either case, that correlation may be characterized by a curve and a measure of goodness of fit - this is an example of a very simple statistical model. Alternatively, mathematical models derive from hypothesized mechanisms / algorithms underlying the dynamics of the studied system. In this case, either algebraic equations or algorithmic instructions determine (a series of) causal relationships among the variables via proposed functional forms (e.g., linear, logarithmic, etc.) to capture the reasons for the relationships. Subsequently, predictions and analysis are made using both the projection of numerical/computation simulations and/or the analytic (i.e., algebraically provable) relationships among the variables as they change according to the rules of the model [18,19,29].

Critically, many models are hybrid models, involving both some mathematical and some statistical techniques, e.g. [66]. For example, to derive the functional forms of the dynamics in a mathematical model, many efforts rely on statistical models to characterize the curves describing the observed patterns in the relationships between the variables. Statistical, mathematical, and hybrid models may all be used in the act of assembling, analyzing, and interpreting data assemblies within and across scales, making them a critical pathway in ensembling [5,42,43,67].

***Complexities of ensembling within statistical and mathematical modeling***

Of the three themes identified within this manuscript, statistical and mathematical modeling has the most nuanced relationship with ensembling processes. As with framework analysis, complexities may arise in how data are interpreted and included within models, the analytic process, and how the results are interpreted. Nevertheless, different elements of information (i.e. data assemblies, results from previous studies) are meaningfully ensembled to reach conclusions about the relationships between phenomena of interest [1,2,20,22]. Each of the complexities discussed previously regarding including and interpreting data assemblies in our framework analysis classification, also apply to statistical and mathematical models.

When the data assemblies used in the modeling process contain the initially observed phenomena of interest, we may directly include them within our statistical model as a straightforward combination of independent variables influencing the dependent variable [32,40]. For example, the rain station team wants to work with a university to explore the conditions that might favor naturally caused wildfires. The team utilizes a multiple regression model, directly including ground saturation and lighting strike data in the model (i.e., independent variables), to explore their influence on the frequency of wildfires in the region. Beyond multiple regression-type analyses, data assemblies may be used to construct more nuanced statistical models that capture multiple variables' synergistic influence. Continuing with the wildfire example, perhaps the team suspects that ground saturation is internally autocorrelated while lightning strikes are stochastic events, but with their own seasonal patterns in frequency of occurrence and uses this extra information to predict a time series for each, to provide a joint Bayesian prior for wildfire over time. That model is then informed by the data in the assemblies, but no longer relies on it directly.

This progression in the use of data assemblies can lead to more complex (and potentially sequentially dependent) statistical models that build to provide more insight than would be possible by direct consideration of the initial observational data. Therefore, the foundational information driving each next model shifts from data assemblies themselves to other models and their results. At the current cutting-edge of ensembled models, dynamic and/or conditional weights may be assigned to the outcomes of a set of independent models that then jointly allow for more accurate predictions than could any individual model alone. For example, cutting-edge efforts in the international climate science and (separately) epidemiological forecasting communities have begun to assess the performance of such sets of models, assembled into a single projection [68–71]. "Doubly- or triply-robust" techniques using multiple different models to control for confounding in observational studies – such as the simultaneous use of traditional regression adjustment as well as propensity scores or inverse-probability weights – can also be viewed as an ensembling method in this light. In this case, the goal of the relatively small ensemble of models is to attempt to ensure the ensemble fully adjusts for any confounding relationship between an exposure, outcome and attendant covariates, where a single model might fail to do so [18].

**Conclusions**

As the frequency of "ensemble" and ensemble-related processes in the literature grows, including a greater diversity of disciplines and application areas, the associated lexicon continues to increase in complexity. Rather than provide an exhaustive index of terms for ensemble phenomena across all existing applications or push for the adoption of a single, universal ensemble-related language, our goal is to provide a unifying concept map capturing the diverse set of methods and logic that involve ensembling of data, data sets, data repositories, and models (Fig 4).

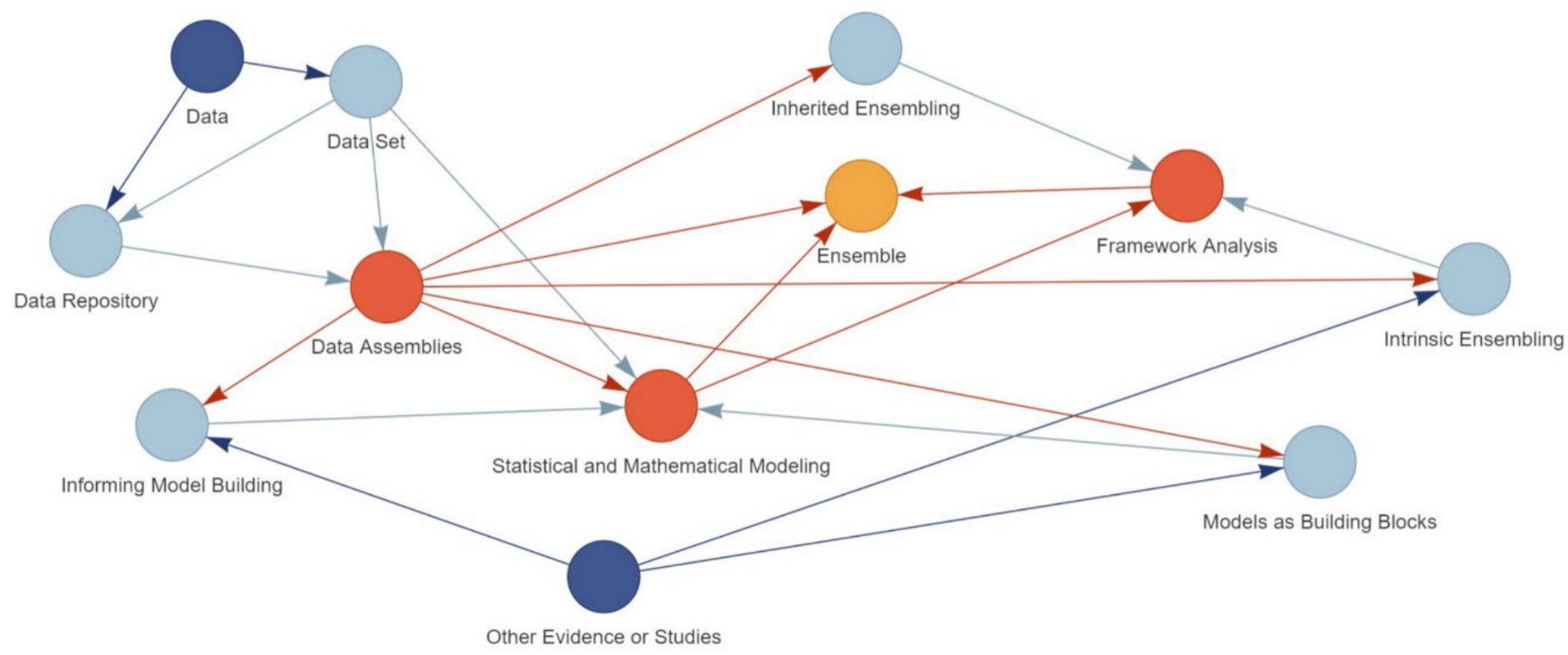

**Fig 4. A visualization of the concept map laid out in this manuscript.** The unifying concept map depicting our perspective of the diverse set of methods and logic related to ensembling in existing academic literature.

This piece does not capture every instance or discipline in which ensemble-related processes occur. Nevertheless, the included living compendium (Fig 2) is a platform for the continued expansion of the disciplines and applications to be included for consideration, and a logic-driven concept map (Box 1) of the scope of ensembles and ensemble modeling. Ultimately, we hope this work will facilitate the continued discovery and advancement of ensembling techniques across and within different disciplines.

**Box 1.** An annotated typology of our concept of the scope of ensembles and ensemble modeling. *Italics are invoked when the text is meant to explain or clarify the point*

1. Measuring One Thing
    a. Individual elements of data are assembled into a data set
    b. Data come from multiple sources, but all sources are still measuring the same thing
        i. Measured in the same way
        *For example, random forest models - ensemble of decision tree models essentially meant to boost accuracy and prevent overfitting - if you are getting a measurement out of the model*
        *Also, meta-analysis wherein the studies being combined have a shared outcome and analytical strategy (i.e. a consortium of RCTs with shared statistical plan).*
        ii. Not measured in the same way
        *Also meta-analysis, probably in its more common form, which is to try to aggregate different study types with the same outcome but different adjustment sets into a pooled estimate*
        iii. Initial measurements from multiple sources are used to generate a single statistical model that can be used instead of observational data thereafter

2. Measuring Many Things
    a. How do you get the data?
        i. All measurements of all things come from the same source
        *For example, when we think of environmentally driven time-series dynamics of vector-borne diseases, the latent variable 'climate' may in fact be meteorological variables such as precipitation, minimum and maximum temperature, average wind speed, relative humidity, which can all be collected in one weather station, reported at the same frequency (e.g. daily), and provide multiple input time-series for data-fitting models.*
        ii. Measurements come from multiple sources
            1. Each thing measured is measured only by one source

*Such as rainfall for a particular area at a rain station - can be many rainstations, set up by WMO with the same frequency mandate for reporting*

        2. Measurements of each thing are from different sources
            a. Measured in the same way, no matter the source
            *Rainfall may be measured by a citizen science initiative that provides the same graduated rain collectors to all participants with clear instructions for their use and reporting*
            b. Not measured in the same way
            *An LLM is used to scrub newspaper-based meteorological reporting of rainfall in multiple cities over time*
    iii. Initial measurements generated a statistical model that can be used instead of observational data thereafter
b. What do we think the data mean?
    i. The things measured are independent of each other
        1. The things measured are each imperfect proxies of the desired outcome
        *In predicting the responses of herbivores to a landscape, one might try to approximate vegetation biomass via a greenness index measured with reflectance, or assume that there are primary productivity responses to rainfall/temp patterns in a proxy called 'phenology'.*
            a. One measure would be enough/best if it were accurately measurable
            b. Best measure is a combination of independent factors because they are imperfect in different ways
                i. Protocol is to switch among them to use one at a time, but only in conditions under which that one is best
                ii. Protocol is to combine them as a mixture, doing better than any one of them
                    *1. Note: If this is fully independently dynamic, it can reduce to the “switch to the best” case*

2. If this is dynamic, but it is unknown how to define or discover which is “best”, it is a case of “how should each component be weighted now” (*in ways that are not dependent on the values of the other measured things*)
    2. The things measured are each component drivers of the desired outcome, but do not impact each other
        a. They all have consistent impact on the outcome
            i. Protocol is to find the best set of things to use a constant set
        b. The relative impact each has on the outcome shifts (but not in a way that depends on any of the others)
            i. This becomes “how do we switch dynamically among weighting different stable sets of things”
ii. They are dependent on each other
    1. Things are correlatively interdependent *(i.e., they are all different proxies of the same thing)*
        a. Basically different ways to measure the same thing *(i.e. redundant)*
        b. With different imperfections in measurement
        c. With different functional forms of relation to the outcome
            i. With similar imperfections
            ii. With different imperfections
    2. The things influence each other as a causal path on the way to impacting the outcome
        a. Subsets can be measured and still provide insight
            *i.e., if measured things M, N, and O all impact the outcome X because M impacts N which impacts O which impacts X directly, it may be enough to observe only M, N, or O and the other two provide no additional information*
        b. Subsets of things being measured do not provide insight

*i.e. M, N and O all contribute to the outcome directly, but how much each contributes depends on the contribution of the 2 others – all three need to be measured to understand how they impact the outcome*

c. How do we understand the things (*these can be in combination with each other*)
    i. Each one is just a direct measurement and we build a single statistical model of all of them as a straightforward combination of independent variables influencing the dependent variable

        *i.e. multiple regression or PCA*

        *Appropriate, for example, for 2.b.i.2.a*

    ii. Each one is a measurement and we build a non-trivial statistical model of all of them

        *e.g., multiscale impacts, lag, etc.*

    iii. Some or all things are included by a statistical model of just that thing
        1. We have a combination of models rather than a combination of data itself
        2. Following the branching from 2.b.i but now for "models" instead of "things"
    iv. Some or all things are included by a statistical model that includes conditional dependencies
        1. Follow the branching from 2.b.ii but now for "models" instead of "things"
    v. Some or all things are included by a mathematical model that posits causal influences via functional forms
        1. Things can be measured data, statistical models, or hypothesized relational functions

            *Note; If it's just measured data directly, it probably won't be called anything like an ensemble model*

3. Putting Together Models
    a. Already in 2.c we discussed the possibility of models of different (sets of) things
    b. Outputs of sets of models are treated as things
        i. i.e., following branching from 2.b for "sets of models"

*As a few further examples (in addition to examples in the main text):*

*Ecological niche model output ensembles, in which the outputs (rather than the inputs) are ensembled - Suppose the authors conducted multiple types of niche models (SDMs, ENMs, etc.) with the same data and environmental parameter inputs and calculate the average outcomes. They then generate several model outputs (using different algorithms) as suitability on gridded data, and use a thresholding to create binary 1/0 suitability outcomes, then weight how many of the models 'agree' in a given pixel.*
*Medicine - Models for predicting the impact of different treatments on individual health outcomes (personalized medicine).*
*Environmental health - Models for predicting the dispersion of pathogens (anthrax, influenza, etc.) in different settings.*

Note: The Assembly Does Not Have to Include Components from Only within the Same Scale
*Data from different sources can lead to one simple multivariable regression, which is then one model in a set of models that are weighted depending on system condition and then the outcome of that set of models combined under weighting can be averaged with a causal model on inferred, unobservable processes that experts hypothesize.*

**Acknowledgments**

We are thankful for our funding, NSF CCF Award 2200140, NSF DBI Award 2412115, and to GH for her help with figure concept design.

## Data reporting

Not applicable.

## Data availability statement

All data and code used within this manuscript are publicly available at https://github.com/bleicham/Assembling-Ensembles and https://doi.org/10.5061/dryad.8gtht771c. The web-based dashboard produced by the code and data provided in the aforementioned links can be found at https://collabdashboard.shinyapps.io/Shiny-Dashboard/#section-about.

## Financial Disclosure Statement

The funders had no role in study design, data collection and analysis, decision to publish, or preparation of the manuscript.

## Competing Interests

The authors have declared that no competing interests exist.

## Author Contributions

Conceptualization: NHF, ETL, JMR, SJR, AB; Investigation: NHF, JMR, SJR, AB, ETL; Writing – Original Draft Preparation: NHF, AB, JMR, SJR, ETL, LB, GC; Writing – Review & Editing: NHF, AB, JMR, SJR, ETL, LB, GC